\documentclass{aa}
\usepackage{graphicx}
\begin{document}

\title{Apparent radii of neutron stars
  and equation of state of dense matter}

\author{P. Haensel\inst{1}}
\institute{N. Copernicus Astronomical Center, Polish
       Academy of Sciences, Bartycka 18, PL-00-716 Warszawa, Poland\\
~~e-mail: {\tt  haensel@camk.edu.pl
  }}
\offprints{P. Haensel}
\date{ 
Received  20  July 2001/Accepted 20 September 2001 
}
\abstract{
Apparent (radiation) radius of neutron star,
$R_\infty$, depends on the star gravitational mass
in quite a different way than the standard coordinate
radius in the Schwarzschild metric, $R$.
We show that, for a broad set of equations
of state of dense matter,
$R_\infty(M_{\rm max})$
for the configurations with maximum allowable masses 
is very close to 
 the absolute lower bound on $R_\infty$ at fixed
$M$, 
 resulting from the very definition of
$R_\infty$. 
Also, the value of $R_\infty$ at given $M$, 
 corresponding to the
maximum compactness (minimum $R$) of neutron star consistent with
general relativity and condition $v_{\rm sound}<c$, 
  is only 0.6\%  higher than this absolute 
lower bound. 
Theoretical predictions for $R_\infty$ are compared with 
existing observational estimates of the apparent radii of 
 neutron stars.
\keywords{dense matter -- equation
 of state -- stars: neutron -- stars}
}
\titlerunning{ Apparent radii of neutron stars and equation of state }
\authorrunning{P. Haensel}
\maketitle
\section{Introduction}
%
Measuring the spectrum of
radiation from neutron star surface (or, more
precisely, atmospheres), combined with knowledge
of distance, 
enables one, assuming spherical symmetry, 
 to determine total luminosity,
effective surface temperature, and neutron star
radius. Recently, such studies have been carried out
 for Geminga (Golden \& Shearer 1999) and RX J185635-3754
(Walter 2001); distance from these isolated neutron stars 
was obtained from measuring of the annual parallax.
 Very recently, {\it Chandra} observations of X-ray sources 
in the globular clusters (whose distance is known 
with relatively good precision)  were proposed  
(and applied) to  calculate 
the radius of a neutron star in quiescence (Rutledge et al. 2001). 

Neutron star are relativistic objects, and for masses
above solar mass their radii may be  only 1.5 - 2  times larger
than the gravitational (Schwarzschild) radius
$r_{\rm g}\equiv 2GM/c^2= 2.95~(M/{\rm M}_\odot)~$km.
Therefore, because of a sizable spacetime curvature
close to neutron star, one has to distinguish between
the ``true'' or ``coordinate'' radius, $R$, which is the radial
coordinate of the stellar surface in the Schwarzschild
metric, and the ``apparent radius'' (sometimes called
``radiation radius''), $R_\infty$,
as determined by a distant observer
studying radiation from neutron star surface.

In the present letter we calculate dependence of $R_\infty$
on neutron star mass for a broad set of equations of state
of dense matter. We discuss  properties 
of the theoretical 
 $R_\infty(M)$ curves,
and finally we confront theoretical calculations
with  recent observational 
estimates of apparent radii of neutron stars. 

\begin{figure}
\resizebox{\hsize}{!}{\includegraphics{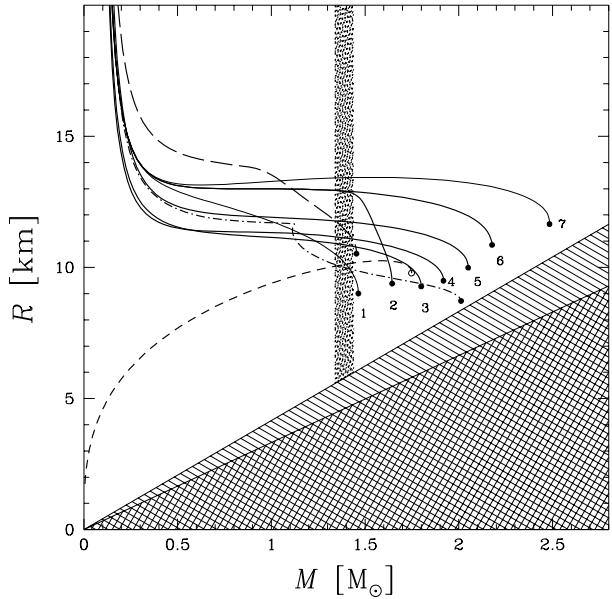}}
\caption{
Neutron star radius $R$ versus gravitational mass $M$,
for seven EOS of baryonic matter, labeled by
numbers 1-7.   1:  BPAL12 of Bombaci
et al. (1995); 2:  EoSN1H1 of Balberg et al. (1999);
3: FPS of Pandharipande \& Ravenhall (1989); 4: Baldo
et al. (1997); 5: Douchin \& Haensel (2000);
6: EoSN1 of Balberg et al. (1999); 7:
EoSN2 of Balberg et al. (1999). Dotted line
corresponds to strange stars built of self-bound
quark matter (SQ1, Haensel et al.1986). Long dashes: hybrid 
neutron stars of dense matter with a mixed baryon-quark phase, 
EOS from Table 9.1 of Glendenning (1997). 
Long dashes-dot line: EOS with first-order phase transition 
to a pure kaon-condensed matter (Kubis 2001). Doubly hatched
area is prohibited by general relativity and
corresponds to $R<{9\over 8}r_{\rm g}$. Singly
hatched area is excluded by general relativity
combined with condition $v_{\rm sound}<c$. In the
case of stars built of baryonic matter, configurations
with maximum allowable mass is indicated by a filled circle,
 and in the case of strange stars,
built of
self-bound quark matter  - by an open circle.
Shaded vertical band corresponds to the range of precisely
measured
masses of binary radio pulsars.
  }
 \label{RM}
\end{figure}

\begin{figure}
\resizebox{\hsize}{!}{\includegraphics{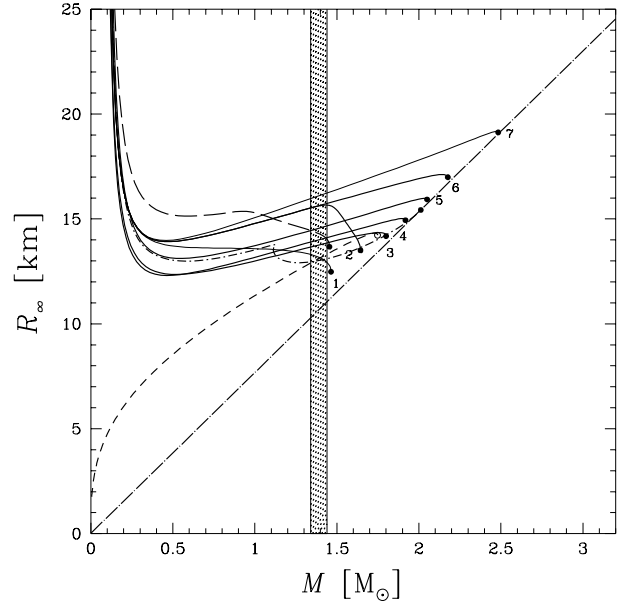}}
\caption{
Apparent radius of neutron star, $R_\infty$,
versus gravitational mass, $M$. Notation as
in Fig. 1. Thick long-dash - dot straight line corresponds to
minimum $R_\infty$ at a given $M$. 
  }
 \label{RMinf}
\end{figure}
\section{$R(M)$, $R_\infty(M)$, and their theoretical
lower bounds}
The effective surface temperature, $T_{\rm s}$, at the
neutron star surface,  is related to total photon
luminosity, $L_\gamma$, by
%
\begin{equation}
L_\gamma=4\pi R^2 \sigma_{\rm SB} T_{\rm s}^4~,
\label{Lgamma.surf}
\end{equation}
%
where all quantities are measured by a local observer on
neutron star surface. Spherical symmetry is assumed.
A distant observer (``at infinity'')
will measure  ``apparent luminosity'' $L^\infty_\gamma$, ,
``apparent effective temperature''  $T^\infty_{\rm s}$, and
``apparent radius'' $R_\infty$, related to quantities
appearing in Eq.(\ref{Lgamma.surf}) by
 (Thorne 1977)
%
\begin{eqnarray}
L_\gamma^\infty = L_\gamma\left(1-{r_{\rm g}\over R}\right)=
4\pi R_\infty^2 \sigma_{\rm SB} \left(T_{\rm s}^\infty\right)^4~,
\cr\cr
T_{\rm s}^\infty=T_{\rm s}\sqrt{1- {r_{\rm g}\over R}}~,
~~~R_\infty = {R\over \sqrt{1- {r_{\rm g}/ R}}}~.
\label{infty-surf}
\end{eqnarray}
%
As we will show, dependence of $R_\infty$ on neutron star mass differs
considerably from $R(M)$; the difference, which reflects spacetime
curvature near neutron star, increases with increasing $M$, and becomes
quite large at the maximum allowable mass, $M_{\rm max}$.
The curves $R(M)$ and $R_\infty(M)$, calculated for a broad
set of equations of state (EOS) of dense matter, are presented
in Fig. 1 and Fig. 2, respectively. One notices, that for
moderately stiff and stiff equations of state ($M_{\rm max}\ga
1.8~{\rm M}_\odot$) without a strong softening at highest densities, 
for $M\ga 0.5~{\rm M}_\odot$
 the apparent radius 
$R_\infty$ {\it increases} with  increasing $M$ (except
for a tiny region close to $M_{\rm max}$), in contrast to
$R(M)$, which {\it decreases}  in the same mass interval.

In Fig. 1, straight lines, marking upper boundaries of the
hatched regions of the $R-M$ plane, result from quite
general physical conditions imposed on the configurations
of hydrostatic equilibrium in general relativity.
The lower boundary results from the condition that pressure
within an equilibrium configuration should be finite,
and can be expressed as $R(M)>{9\over 8}r_{\rm g}$
(Buchdahl 1959; general proof can be found in Weinberg 1972).
 This condition  can rewritten as
$R(M)>R_{\rm min1}(M)=3.32~(M/{\rm M}_\odot)~$km.
A stronger condition is obtained if  we additionally
require that sound speed within the star should
be subluminal: $v_{\rm sound}=\sqrt{{\rm d}P/{\rm d}
 \rho}<c$ (such a condition is {\it necessary}, 
but {\it not sufficient} (Olson 2000),   to respect
causality in a fluid medium).
The condition  $v_{\rm sound}<c$ implies a lower 
bound on $R$ at a given $M$ (Lindblom 1983, Haensel \& Zdunik 
1989, Lattimer et al. 1990, Glendenning 1992, Haensel et al. 1999, 
Glendenning 2000). 
In what follows, we will use numerical values of the 
{\it absolute} lower bound 
$R_{\rm min2}(M)=4.17~(M/{\rm M}_\odot)~$km, as calculated 
in Haensel et al. (1999); the older values obtained in 
(Lattimer et al. 1990, Glendenning 1992) are slightly higher, 
because of the assumed presence of an outer envelope (crust) 
in neutron star models, while the value deduced from (Lindblom 
1983) is not very precise (see Haensel et al. 1999). 

A strict lower bound on $R_\infty(M)$ results from the
very definition of $R_\infty$ (Lattimer  \&  Prakash 2001). 
Namely, the definition of $R_\infty$ implies
%
\begin{equation}
{R_\infty\over r_{\rm g}} =
{R\over r_{\rm g}\sqrt{1- r_{\rm g}/R}}~.
\label{Rinf.rg}
\end{equation}
%
The right-hand-side of the above equation is a function of
$x\equiv r_{\rm g}/R$ only.
 It diverges to $+\infty$ at $x
=0$ and at $x=1$. At fixed $M$, it has a single minimum
at $x=2/3$. Therefore, minimum value of $R_\infty(M)$
is $R_{\rm \infty,min}(M)=7.66~(M/{\rm M}_\odot)~$km
(Lattimer \& Prakash 2001).
 
Let us notice, that this  limiting
$R_\infty$ for an  ``apparently
most compact'' neutron star is very close to
(but a little smaller than)
that for a maximum compactness $x=0.7081$
consistent with $v_{\rm sound}<c$, 
given by $7.71~(M/{\rm M}_\odot)$~km. However, at
any $M$ the difference is only 0.6\%, and therefore in
practice smallest $R$ at a fixed $M$, consistent with
$v_{\rm sound}<c$, can be considered as corresponding  to
smallest $R_\infty(M)$, and vice versa.

This result can be easily understood, because
$x=0.7081$ is only by 0.05 higher than 2/3 corresponding
to the minimum of $R_\infty$. Therefore, relative difference
between $ R_{\rm \infty,min}$ and the value corresponding
to minimum value of $R$ at given $M$  (assuming
$v_{\rm sound}<c$), can be estimated as 
  $\simeq 2.25\times (0.05)^2 = 0.6\%$, which is
consistent with our exact result.

While the subluminal ($v_{\rm sound}<c$) upper  bound on $x$ at
given $M$ is slightly larger than 2/3, the {\it actual} maximum
values of $x$ for various EOS, which are  reached
at $M_{\rm max}$ for these EOS, are lower than 2/3. However,
if we restrict to medium stiff and stiff EOS, with $M_{\rm max}
\ga 1.8~{\rm M}_\odot$, then $x_{M_{\rm max}}
\simeq 0.6$, which is only 0.06 lower than 2/3. We may therefore
expect, that for these EOS, which actually constitute majority
of models in Figs.1, 2,  $(R_\infty)_{M_{\rm max}}$
will be only $\sim 2.25\times (0.06)^2 \simeq 
0.8\%$ larger than
$R_{\rm \infty,min}$. This explains, why for these EOS the
points at $M_{\rm max}$ are so close to the $R_{\rm \infty,min}(M)$
line in Fig. 2. 

As shown by Lattimer \& Prakash (2001), one expects that 
for any baryonic 
 baryonic EOS, $R_\infty > 11$~km,
independently of neutron star mass. Our Fig. 2 confirms 
this ``practical lower bound''  on $R_\infty$. 
On the contrary, there is
no lower limit on $R_\infty$ for bare strange stars, whose 
size can be as small as hundred fermis. For strange stars covered 
with a layer of normal matter, minimum radius is reached at 
$M\sim 0.01~{\rm M}_\odot$, and for a maximally thick crust it is 
about $5-6~$km (see, e.g., Glendenning 1997). 
\section{Observational bounds on $R_\infty$}
%
In what follows we will briefly review  observational determination 
of apparent radius of neutron star. We will restrict to  cases, 
which seem to us most promising. In all cases, what one determines 
is actually  $R_\infty/d$, where $d$ is neutron star distance. 
Therefore, independent knowledge of  $d$ is mandatory to calculate 
$R_\infty$ from observational data. Generally, after fitting  
the spectrum of photons emitted  from neutron star, one tries 
to get the interval  $R_{\infty,{\rm l}}<R_\infty
<R_{\infty{,\rm u}}$, to which the value $R_\infty$ 
belongs at not less than 90\% confidence level. 
An EOS is considered to be ruled out, 
if no point on its  $R_\infty(M)$ curve 
can satify this condition. In practice, the condition reduces 
to $R_\infty<R_{\infty,{\rm u}}$. Generally, conclusions 
from the application of this criterion should be taken with 
a grain of salt, because of the difficulty in estimating 
of the error in the photon spectrum fitting.  
\subsection{Close by isolated neutron stars}
At  the $95\%$ confidence level, 
results of Golden \& Shearer (1999) imply 
 $R_\infty<R_{\infty,{\rm u}}=
17.6$~km assuming the H atmosphere, and $R_\infty
<R_{\infty,{\rm u}}=16.5$~km
for the black body thermal spectrum (which turns out to be
practically indistinguishable  from the
Fe/Si model atmosphere spectrum). Therefore, 
the value of the upper bound
on the apparent radius of Geminga
is, fortunately, 
 not very sensitive to the assumed atmospheric model.

As for this writing, the case of RX J185635-3754 (Walter 2001, and references 
therein), is much less clear. 
Using atmosphere  model of the photon spectrum, one deduces 
from numbers quoted in Walter (2001) the upper bound  
 $R_{\infty,{\rm u}}({\rm atm})=
18$ km
at the $2\sigma $ confidence level. However, 
 if  one uses the black body spectrum model, assuming 
spherical symmetry, one gets  
 an abnormally small upper bound,  
  $R_{\infty,{\rm u}}({\rm BB})=8.4$~km (Walter 2001; notice 
that we use results  at the $2\sigma$ confidence level). 
Clearly, these results are very preliminary, and we have still to wait 
 for more reliable and less model dependent determinations of $R_\infty$ 
for this isolated neutron star.
\subsection{X-ray transients in globular clusters}
Very recently, Rutledge et al. (2001) proposed a method  
of measuring $R_\infty$ of neutron stars, observed as 
 X-ray transients in globular clusters.  
As an example, they 
studied 
transcient X-ray source 
CXOU 132619.7-472910.8 in 
NGC 5139, 
fitting photon spectrum with the 
H-atmosphere model. They 
obtained, at $90\%$ confidence level, 
$R_\infty = 14.3\pm 2.5~$km (assuming 10\% uncertainty 
in the distance to NGC 5139), which results 
in $R_{\infty,{\rm u}}=
16.8~$km. 
The advantage of the proposed method stems from the fact, 
that for neutron stars located in globular clusters both 
the distance and interstellar  hydrogen column density 
are rather well known. 
\subsection{Other estimates of neutron star radii}
They are mostly related to accreting neutron stars, observed as 
X-ray bursters (see, e.g., Titarchuk 1994, Haberl \& Titarchuk 1995, 
Burderi \& King 1998, Psaltis \& Chakrabarty 1999, Li et al. 
1999a, Li et al. 1999b, and references therein). One has to 
mention a strong model dependence of theoretical analyses, and frequent 
neglect (Burderi \& King 1998, 
 Psaltis \& Chakrabarty 1999, Li et al. 
1999a, Li et al. 1999b) 
of the space-time curvature effects, which are 
actually crucial 
for the difference beteen $R$ and $R_\infty$. 
\subsection{Theory versus observations}
As one can see in Fig. 2, 
the  upper bounds 
$R_{\infty,{\rm u}}$ 
on the apparent radius  of Geminga 
are consistent with theoretical
predictions for $R_\infty(M)$, 
 based on
 considered baryonic EOS of dense matter, 
 provided the neutron star 
mass  is above $0.2~{\rm M}_\odot$.

The value of the upper  bound $R_{\infty,{\rm u}}=16.8~$km, 
obtained by Rutledge et al. (2001) for the 
transient X-ray source 
CXOU 132619.7-472910.8 in 
NGC 5139, is consistent with neutron star $R_\infty(M)$ curves 
 in Fig. 2, 
provided  $M>0.4~{\rm M}_\odot$. In the case of stiffest EOS 
(with $M_{\rm max}>2.2~{\rm M}_\odot$), the value of 
$R_{\infty,{\rm u}}$ rules out high-mass neutron stars 
with $M>1.6\div 2.0~{\rm M}_\odot$.

As for this writing, attempts to estimate the apparent 
radius of RX J185635-3754 (Walter 2001) 
are very model dependent. 
Nevertheless, we are tempted to 
  make a following 
comment.  
Had we accepted the estimate $R_\infty({\rm BB})$ for 
 RX J185635-3754, 
this object could be but a low-mass strange star. In order 
to produce thermal photon spectrum, this low-mass strange star should 
have been covered with a layer of normal matter, because a bare quark surface 
would be a too weak photon emitter (Chmaj et al. 1991, Usov 2001). 
One can only hope that more precise measurement of the photon spectrum 
for this isolated neutron star will liberate us from such basic 
ambiguities.
\section{Summary}
%
 Detection of photons emitted from the surface of isolated neutron stars of known
 distance can result in determination of the apparent neutron star radius,
 $R_\infty$. Due to significant space-time curvature, dependence of $R_\infty$ on
 on stellar mass is quite different from that of the standard ``coordinate'' radius
 $R$. The very definition of $R_\infty$ implies a lower bound, obtained 
by Lattimer \& Prakash (2001), of  $7.66~(M/{\rm
 M_\odot})~$km. 
  At any $M$, this lower limit is very close
 to the value of $R_\infty$ corresponding  to the minimum  $R$, calculated under
 the condition of subluminal sound. Simultaneously, the actual 
values of $R_\infty$
 calculated at the maximum allowable mass are also  close to this limit. For
 moderately stiff and stiff EOS with $M_{\rm max}\ga 1.8~{\rm M}_\odot$, the actual
 value of $R_\infty(M_{\rm max})$ is less than one percent higher than the absolute
 lower bound on $R_\infty$ at this value of stellar mass.

Most reliable  observational estimates of $R_\infty$, obtained 
for Geminga  and transient X-ray source 
CXOU 132619.7-472910.8 in 
NGC 5139, 
  lead to upper bounds on $R_\infty$, 
which are consistent  with all considered baryonic EOS, provided the mass 
of neutron star is above $0.2\div 0.4~{\rm M}_\odot$. 
 
\begin{acknowledgements}
I am grateful to J.L. Zdunik for the reading of the manuscript and for 
helpful remarks. I am  also grateful to A. Potekhin for his precious help in
the preparation of figures.
This research
was partially supported by the KBN grant No. 5P03D.020.20.
\end{acknowledgements}

\end{document}